%% file: paper_1.tex
\documentclass[fleqn,usenatbib]{mnras}
\usepackage{newtxtext,newtxmath,tikz}
\usepackage[utf8]{inputenc}
\usepackage[T1]{fontenc}
\usepackage{ae,aecompl}
\usepackage{graphicx}
\usepackage{amsmath}
\usepackage{amssymb}
\usepackage{xcolor}
\usepackage{soul}
\definecolor{nicegreen}{HTML}{2CA02C}
\definecolor{darkgreen}{rgb}{0.0,0.5,0.0}

\newcommand{\proba}{\mathcal{P}}
\newcommand{\mvec}[1]{\boldsymbol{#1}}
\newcommand{\mmat}[1]{\mathbf{#1}}
\newcommand{\Msun}{$h^{-1}~\mathrm{M}_\odot$}

\usetikzlibrary{shapes}

\newcommand{\borg}{\textsc{borg}}

\title[Neural bias model]{Neural physical engines for inferring the halo mass distribution function}
\author[T. Charnock et al.]{Tom Charnock$^{1}$\thanks{E-mail: charnock@iap.fr (IAP)}, Guilhem Lavaux$^{1, 2}$, Benjamin D. Wandelt$^{1, 2, 3}$, \newauthor Supranta Sarma Boruah$^{4,5}$, Jens Jasche$^{6}$ and Michael J. Hudson$^{7,5,8}$\\
$^{1}$Sorbonne Universit\'e, CNRS, UMR 7095, Institut d'Astrophysique de Paris, 98 bis boulevard Arago, 75014 Paris, France\\
$^{2}$Sorbonne Universit\'es, Institut Lagrange de Paris, 98 bis boulevard Arago, 75014 Paris, France\\
$^{3}$Center for Computational Astrophysics, Flatiron Institute, 162 5th Avenue, 10010, New York, NY, USA\\
$^{4}$Dept.\ of Applied Mathematics, University of Waterloo, 200 University Avenue West, Waterloo, ON, Canada  N2L 3G1\\
$^{5}$Waterloo Centre for Astrophysics, University of Waterloo, 200 University Avenue West, Waterloo, ON N2L 3G1 Canada\\
$^{6}$The Oskar Klein Centre, Department of Physics, Stockholm University, Albanova University Center, SE 106 91 Stockholm, Sweden\\
$^{7}$Dept.\ of Physics and Astronomy, University of Waterloo, 200 University Avenue West, Waterloo, ON N2L 3G1 Canada\\
$^{8}$Perimeter Institute for Theoretical Physics, 31 Caroline St N, Waterloo, ON N2L 2Y5, Canada
}

\date{Accepted XXX. Received YYY; in original form ZZZ}
\pubyear{2019}

\begin{document}
\label{firstpage}
\pagerange{\pageref{firstpage}--\pageref{lastpage}}
\maketitle

\begin{abstract}
    An ambitious goal in cosmology is to forward-model the observed distribution of galaxies in the nearby Universe today from the initial conditions of large-scale structures. For practical reasons, the spatial resolution at which this can be done is necessarily limited. Consequently, one needs a mapping between the density of dark matter averaged over $\sim$ Mpc scales, and the distribution of dark matter halos (used as a proxy for galaxies) in the same region. 
    Here we demonstrate a method for determining the halo mass distribution function by learning the tracer bias between density fields and halo catalogues using a \emph{neural bias model}.
    The method is based on the Bayesian analysis of simple, physically motivated, neural network-like architectures, which we denote as neural physical engines, and neural density estimation. 
    As a result, we are able to sample the initial phases of the dark matter density field whilst inferring the parameters describing the halo mass distribution function, providing a fully Bayesian interpretation of both the initial dark matter density distribution and the neural bias model. 
    We successfully run an upgraded \borg{} inference using our new likelihood and neural bias model with halo catalogues derived from full $N$-body simulations. 
    We notice orders of magnitude improvement in modelling compared to classical biasing techniques.
\end{abstract}


\section{Introduction}
    Observations of the large scale structure of the universe provide a window that allows us to constrain our physical models of the universe.
    Although cosmological models can predict the statistical nature of the structures that we see, it is difficult to extract the wealth of information that we observe in real objects.
    One is able to make use of these real structures by constraining the initial phases of the dark matter distribution conditional on the observed data.
    This ambitious task is made possible using algorithms such as the Bayesian Origin Reconstruction from Galaxies (\borg{}) algorithm~\citep{Jasche:2013, Jasche:2015, Lavaux:2016}.
    With the \borg{} algorithm, the initial distribution of dark matter is evolved forward to the dark matter density today, at which point a bias model is used to compare to the observed data via a choice of metric (the likelihood).
    In essence, the bias model contains a phenomenological description of the complex astrophysics which dictates how visible observations trace the dark matter distribution~\citep{Peebles:1980, Kaiser:1984}.
    The bias model, therefore, is a parameterised surrogate for the extremely non-linear, scale-dependent, and environment-dependent astrophysics of galaxy formation and evolution. 
    Since bias has finite reach in spatial scales to ensure causality constraints and distinct from the larger-scale effects that imprints cosmological information on the data, one can still gain information about cosmology.
    However, a poor choice of bias model can massively impact the inference of the initial phases due to the mapping between the dark matter distribution and the observables being incorrect~\citep{Elsner:2019}.
    
    \medskip
    In this work, we present a novel suite of methods for learning a bias model based on physically motivated, neural network-like algorithms (neural physical engines, NPEs) and neural density estimators: a neural bias model.
    Furthermore, the parameters of the network will be inferred as part of the feed forward inference of the initial density phases within the \borg{} framework.
    As such, the parameters for such a neural bias model become part of the fully Bayesian interpretation of the constraints.
    By using the physically motivated architectures of a neural physical engine, we can both massively decrease the number of parameters in the model and drastically increase the interpretability of where the information about the bias model arises in the data.
    
    \medskip
    The combination of the neural physical engine and the Bayesian sampling of the parameters of the model provides us with a method for using neural networks \textit{without any training data} since the distribution of weights (parameters) are conditional on the true (observed) data. Our approach is therefore the realisation of an ultimate version of \textit{zero shot learning}; the neural bias model is learned directly from the data to be analysed, without reference to any training data.
    
    \medskip
    For convenience we will limit ourselves to modelling the relation between the large scale dark matter distribution and its embedded small scale halos. 
    Halos are gravitationally bound objects that host the galaxies that we see.
    It is useful to be able to describe the distribution of halos of a particular mass within a given density environment. 
    By learning to parameterise this distribution as a function of the local density field, it becomes possible to sample realisations of the observed halos which can then be constrained using halo catalogues.
    Therefore, this method provides a way to emulate some of the features of the halo-occupation-distribution (HOD) model using only differentiable operations.

    \medskip
    The paper will follow as such. 
    In Section~\ref{sec:HMDF} we will describe the properties of the halo mass distribution function which relates the dark matter distribution today to (observed) halos.
    We will then show in Section~\ref{sec:NBM} that we can build a neural network capable of learning this function from data, and how this network can be made efficient and interpretable using physical principles. 
    We will also explain how the parameters of this neural physical engine can be inferred to provide us with a fully Bayesian interpretation of the neural bias model.
    In Section~\ref{sec:BORG} we will elucidate the \borg{} framework and how the neural bias model can be included.
    And finally, in Section~\ref{sec:results} we present the data simulation, model and results before concluding in Section~\ref{sec:conclusions}.

\section{Halo mass distribution function}
\label{sec:HMDF}
    The halo mass distribution function, $n(M|\delta)$, is a measure of the number of counts of compact objects with a certain mass, given the environment \citep{Press:1974, Bardeen:1986, Mo:1996, Sheth:2004a}.
    In particular, there are a number of different effects which influence the form of this function including the formalism of~\cite{Press:1974} in which the formation of gravitationally collapsed objects from the initial density fluctuations is described by a power-law for small masses with an evolving exponential mass cutoff.
    Further, the effect of the local density field is known to affect the shape of the function \citep{Kaiser:1984, Bardeen:1986, Sheth:2004b}. 
    Stochastic and higher-order effects are also known to influence the mass distribution of halos, in particular the departure from a Poissonian distribution of the bias~\citep{Kitaura:2014,Saito:2014}. 
    
    \medskip
    To properly account for all of the effects one needs to understand all physical effects, including those not yet realised.
    \citet{Desjacques:2018} showed that by considering all causally selected features along the light cone, the complete expression for the next-to-leading-order galaxy power spectrum and the leading order galaxy bispectrum can be described using only 22 parameters.
    We use this principle to motivate a physical mapping between the dark matter density distribution and the dark matter halos.
    To include all relevant information, and allow freedom to correctly fit the true halo mass distribution, we will consider a neural bias model with the ability to learn about the important structures of the environment which affect the number of halos with different masses, directly from the data, without additional training data.
    In this first outlook into learning such tracer biases, we will consider a Poissonian sampling of the the halo mass distribution function, where we allow some freedom from the Poissonity by non-linearly combining local patches of the density field.
    This can be generalised straight-forwardly, and this will be explored in future work.

    \medskip
    For very narrow mass bins $[M,M+\Delta M]$ we may express the Poisson intensity from the expectancy $\lambda_{i,M}$ of observing $N_{i,M}  $ halos within a particular bin:
    \begin{align}
        \lambda_{i,M} & \equiv \langle N_{i,M} \rangle \nonumber\\
        &= V \int_{M}^{M+\Delta M}\text{d}M\; n_i(M|\{\delta_j\}),
    \end{align}
    with $V$ the volume of a voxel of the grid and $i$, $j$ labelling the voxel. In the above, we note that the halo distribution may be voxel dependent.
    For a Poisson likelihood, for which all mass bins and all voxels are independent, we may write the logarithm of the likelihood $\mathcal{L}$ as
    \begin{equation}
        \mathcal{L} = \sum_{M \in\substack{\text{mass}\\\text{bins}}}~\sum_{i \in \text{voxels}} \left(-\lambda_{i,M} + N^{\text{observed}}_{i,M} \log \lambda_{i,M} \right)\;,
    \end{equation}
    with $N^{\text{observed}}_{i,M}$ the number of halos observed in mass bin $M$ and voxel $i$. 
    The above equation is just the sum of the logarithm of the Poisson probability distribution. 
    For very narrow mass bins $N^\text{observed}_{i,M}$ may take only values that are zero or one. We may thus reorganise the summation as
    \begin{equation}
        \mathcal{L} = \hskip-1.2em\sum_{j \in \text{catalogue}}\hskip-1.2em\log \lambda_{i(j),M(j)} - \hskip-0.7em\sum_{i \in \text{voxels}}\hskip-0.7em\lambda_{i,M}\;,
    \end{equation}
    with $i(j)$ and $M(j)$ the index of the voxel and mass bins corresponding to the halo $j$. 
    This equation leads, for $\Delta M \rightarrow 0$, to the following log-likelihood to quantify how a mass distribution is representative of the number of halos (or galaxies) contained in a catalogue
    \begin{equation}
        \mathcal{L} =\hskip-1.2em \sum_{j\in\textrm{catalogue}}\hskip-1.2em\log\left(n(M_j|\delta_j)\right) - V\hskip-0.7em\sum_{i\in\textrm{voxels}}\hskip-0.1em\int_{M_\textrm{th}}^\infty\hskip-0.3em dM~n(M|\delta_i)\;,\label{eq:likelihood}
    \end{equation}
    where we have discarded the constant induced by the logarithm of the infinitesimal binning $\Delta M$.
    The first term involves calculating the halo mass distribution function for every halo of the catalogue given the density at each halo's position on a grid.
    The second term evaluates the halo mass distribution function at every voxel in the gridded density field and integrates over the mass from some mass threshold $M_\textrm{th}$.

\section{Neural bias model}
    \label{sec:NBM}
    
    We wish to build an automatic method for modelling the halo mass distribution function based on the physical principles which we know are relevant, but which is parameterised as simply as possible to provide us with the ability to effectively sample the parameters of this model.
    The added bonus of such a simple model is the increased interpretability of the function.
    How to build such a physically motivated neural network is described in section~\ref{ssec:NPE}.
    Since we want to build the output of the neural bias model to be the halo mass distribution function, we need to ensure that the outputs are evaluations of this function for halos of a certain mass given a density environment.
    To do so we can consider neural density estimators, and in this case a simple mixture density network is sufficient.
    Mixture density networks are explained in further detail in section~\ref{ssec:MDN}.
    Finally, it is essential that we can infer the parameters of our neural bias model to provide us with a fully Bayesian interpretation of the initial density field and the tunable parameters of the network.
    The method by which we are able to efficiently sample these parameters is described in section~\ref{ssec:HMCLET}.
   
    \subsection{Neural physical engines}
    \label{ssec:NPE}
    We will here use terminology familiar to those in the machine learning and computer science communities \citep[for full details see eg.][]{Goodfellow:2016}.
    Much of the work considered derives from the understanding of neural networks.
    Neural networks are described as a set of algorithms with \emph{trainable} parameters known as weights and biases which map an input to an output.
    The algorithms usually consist of simple operations such as matrix multiplications or convolutions between input and weight vectors.
    The output vectors of these operations are then normally acted upon by a scalar \emph{activation} function, which allows one to add non-linearity to the mapping.
    Finally, many \emph{layers} of these operations can be stacked input-to-output, building a great deal of abstraction between the original input and the final output.
    This provides us with a highly complex, non-linear, and arbitrary function which can be fitted using training data.
    
    \medskip
    It is currently fashionable to build extremely large, deep neural networks capable of huge abstraction from the input to the output.
    However, in physics we have models which can drive the design and conception of the architecture that we wish to use.
    The symmetries of the physical model which describes the data can be used to massively reduce both the amount of data and effort needed to train such a neural physical engine.
    Since a neural physical engine is greatly restricted in the freedom along the directions which we know are constrained by the data, there is far less chance of overfitting.
    Along with this, we obtain a much greater interpretability of what the neural network is doing.
    In this work we will denote the neural physical engine $\mathbb{N}(\boldsymbol{\theta}^\textrm{\scriptsize NPE}):\boldsymbol{\delta}\to\boldsymbol{\psi}$, i.e. the physically motivated mapping with tunable parameters $\boldsymbol{\theta}^\textrm{\scriptsize NPE}$ which takes a patch of the evolved density field $\boldsymbol{\delta}$ to a transformed informative vector, $\boldsymbol{\psi}$, for the halo mass distribution function. 
    Note in this work that we consider the size of $\boldsymbol{\psi}$ to be 1 as it represents the transformed density field in the central voxel.
    
    \medskip
    As described at the start of Section~\ref{sec:HMDF}, we already know that the local density fluctuations are important for describing the halo mass distribution, as is the large scale overdensity.
    We also know that the action of the tracing of the dark matter distribution by halos is both translationally and rotationally invariant, although local distortions can give rise to more optimal environments for the development of halos.
    Furthermore, we know that the bias model is non-linear, all of which we can build into our neural bias model.
    
    \medskip
    We start constructing our model using convolutional kernels which respect symmetries.
    Naturally, the idea of convolutional kernels respects the translational symmetry of the problem.
    As is usual in the machine learning literature, convolutional kernels are matrices where every element is a trainable parameter known as a weight.
    Since we also want to build in rotational invariance and local deformations of rotational invariance we expand the kernels in terms of multipoles.
    This provides us with a hierarchy of convolutional kernels with each multipole describing further deformation from exact rotational symmetry.\footnote{The code for generating multipole kernels in \textsc{TensorFlow} (with available \textsc{keras} module) is available at \url{https://github.com/tomcharnock/multipole_kernels}.}
    In three dimensions (as considered here), the basis of rotational symmetries of the convolutional kernels is spherical harmonics.
    In the $\ell=0$ case, such a kernel is constructed by associating the same weight in the convolutional kernel to each matrix position which is equidistant from the centre.
    Then for $\ell=1$ we have $2\ell+1$ kernels where each weight is associated by distance and by the values of $Y^\ell_m(\theta, \phi)$ at angles $\theta$ and $\phi$ from the centre of the matrix.
    This expansion can be continued to learn further information about deformations away from exactly rotationally invariant environments.
    As an example of the drastic savings in the number of weights that one can obtain by expanding the kernels in terms of multipoles consider a $3^3$ kernel. 
    This kernel would traditionally have 27 independent trainable parameters.
    The $\ell=0$ kernel has only four weights, one at the centre and one shared on every edge, every face and every corner.
    The $\ell=1$ kernel has $(2\ell+1)=3$ output \emph{feature} maps, each kernel containing three independent weights.
    In this case, only nine parameters are needed in the place of 27.
    It should be noted that if the $\ell=2$ kernel is also included, all 27 of the available parameters of a $3^3$ kernel are exhausted.
    As these kernels are fitted, using whichever optimisation procedure is most suitable, information is extracted from the data via these orthogonal pathways.
    By analysing the scale of the weights in each of these kernels it becomes obvious to see where the information in the data is contained. 
    That can be for example in the radially symmetric patches of the data, or in the shear component of the data.
    To summarise, the outputs feature maps, $\boldsymbol{C}^{\ell,m}$, of the symmetry respecting convolutional kernels, $\boldsymbol{K}^{\ell,m}$, takes the form 
    \begin{equation}
        C^{\ell,m}_{j} = \sum_{i=-\kappa/2}^{\kappa/2} K^{\ell,m}_{i} \delta_{j-i}\;.
    \end{equation}
    Here, for simplicity, we have written a single sum representing the $N$ summations for a kernel $\boldsymbol{K}^{\ell,m}\in\mathbb{R}^N$ where $\kappa$ represents the size of each of the $N$ dimensions of $K^{\ell,m}$.
    The values of $K^{\ell,m}_i$ depend on the multipole $(\ell,m)$. 
    Each $\ell$ and $m$ kernel has a different prescription of the number of shared weight elements and the position of those shared elements within the kernel.
    
    \medskip
    The non-linearity of the bias model is included via the application of an activation function to the output of the convolutional kernels. Formally, we thus have
    \begin{equation}
        \boldsymbol{\phi}^{\ell,m} = A(\boldsymbol{C}^{\ell,m})\;,
    \end{equation}
    with $A(x)$ a scalar activation function and $\boldsymbol{\phi}^{\ell,m}$ as the $\ell$ and $m$ components of the activated feature maps.
    The more depth in the neural physical engine, the more non-linear the response from the input patch can be.
    Moreover, deep, activated stacks independent along the orthogonal pathways can provide extremely complex functions which remain, for example, rotationally invariant for $\ell=0$, provided the activation function remains scalar.n This is a particular example, relevant to case under consideration here, because we restrict to $\ell=0$. 
    For broader applications coupling the multipoles would drastically broaden the scope of physical effects that can be modelled.
    
    \medskip
    To ensure that the relevant local information is taken into account, the size of the kernel (or the total combined receptive patch of many layers of convolutions) must be large enough to pull all relevant information from the data.
    This size should generally be known thanks to the physical model.
    However, regardless of this, one can construct large spatial multipole kernels if necessary, since the number of parameters is already massively reduced.
    The parameter values from regions of the kernel that are not important will tune to zero and be interpretable as the distance over which information is relevant for the problem.
    Furthermore, the scales of each of the kernels for different multipoles will indicate on which scales the degree of deformation is most important.
    
    \medskip
    Finally, we choose to use the late time overdensity field as our input such that the neural bias model is provided with the non-local information about the density field via the mean of the distribution.
    We could consider, instead, biasing the initial conditions and transporting that along with the dynamical model forming a Lagrangian neural bias model, which folds in the dark matter history. 
    We will explore this option in future works since it is possible that a sufficiently complex neural bias model will be able to obtain these aspects of the history from the final time step.

    \medskip 
    Using these techniques, we build a well-reasoned model which performs the real space mapping to the quantity of interest whilst remaining free enough to fit the complex function, all the while being interpretable.
    
    \subsection{Mixture density networks}
    \label{ssec:MDN}

    To approximate the halo mass distribution function we make use of a very slight variation of a mixture density network \citep[MDN,][]{Bishop:1994}.
    MDNs use neural networks fitted to predict the normalised weights and parameters of several distributions, such as: the mean and standard deviation of a Gaussian distribution; the event rate of a Poisson distribution; or the shape parameters of a $\beta$-distribution.
    By combining a mixture of distributions, new distributions of many different forms can be built
    \begin{equation}
        \mathcal{P}({\bf x}|{\bf d}, \boldsymbol{\theta}) = \sum_{i=1}^N \alpha_i({\bf d}, \boldsymbol{\theta})\mathcal{F}_i({\bf x}|{\bf d}, \boldsymbol{\theta})
    \end{equation}
    where $\mathcal{P}({\bf x}|{\bf d},\boldsymbol{\theta})$ is the mixture of probability distributions, $\mathcal{F}_i({\bf x}|{\bf d}, \boldsymbol{\theta})$ is the collection of $N$ parameterisable distributions, $\alpha_i({\bf d}, \boldsymbol{\theta})$ is a set of amplitudes for each distribution and ${\bf x}$, ${\bf d}$ and $\boldsymbol{\theta}$ are the parameters of the model to be inferred, the input data and the trainable parameters of the neural network, respectively. 
    
    \medskip
    To ensure that the output of the mixture density network can be interpreted as a probability distribution, the set of $\alpha_i({\bf d}, \boldsymbol{\theta})$ are \emph{activated} using a softmax function, defined as
    \begin{equation}
        \mathrm{softmax}(\boldsymbol{\alpha}) = \frac{\alpha_i}{\sum_{i=1}^N \alpha_i}\;.
    \end{equation}
    This ensures that the amplitudes are normalised.
    In principle, the number of distributions in the MDN are self-regulating since the amplitudes of any irrelevant distributions vanish.
    This self-regularisation can lead to problems with Bayesian sampling of the network since there is strong multi-modality.
    This multi-modality is in practice solved by restricting the parameter space as we show in equation~\eqref{eq:mu}.
    
    \medskip
    We are interested in approximating the number density of halos with a certain mass given a density environment, $n(M|\delta)\equiv \bar{N} \mathcal{P}(M|\boldsymbol{\psi}, \boldsymbol{\theta}^\textrm{\scriptsize MDN}(\{\delta_j\}))$ where $\boldsymbol{\psi}$ is the single-valued vector output of the neural physical engine, $\boldsymbol{\psi}$, discussed in section~\ref{ssec:NPE} and $\bar{N}$ is the mean number density.
    Since the number density is not a probability distribution we do not constrain the set of amplitude parameters $\alpha(\boldsymbol{\psi}, \boldsymbol{\theta}_i^{\boldsymbol{\alpha}})$ using the softmax function, and instead just ensure their positivity.
    Since the halo mass distribution function is relatively smooth (as seen via the diamonds in Figure~\ref{fig:hmdf}) we decide to model it using a mixture of Gaussians
    \begin{align}
        n(M|\delta) & = \sum_i^N\alpha(\boldsymbol{\psi}, \boldsymbol{\theta}_i^{\boldsymbol{\alpha}})\mathcal{N}\left(\mu(\boldsymbol{\psi}, \boldsymbol{\theta}_i^{\boldsymbol{\mu}}), \sigma(\boldsymbol{\psi}, \boldsymbol{\theta}_i^{\boldsymbol{\sigma}})|M\right),\nonumber\\
        & = \sum_i^N\frac{\alpha(\boldsymbol{\psi}, \boldsymbol{\theta}_i^{\boldsymbol{\alpha}})}{\sqrt{2\pi\left(\sigma(\boldsymbol{\psi}, \boldsymbol{\theta}_i^{\boldsymbol{\sigma}})\right)^2}}\exp\left[-\frac{\left(\log(M) - \mu(\boldsymbol{\psi}, \boldsymbol{\theta}_i^{\boldsymbol{\mu}})\right)^2}{2\left(\sigma(\boldsymbol{\psi}, \boldsymbol{\theta}_i^{\boldsymbol{\sigma}})\right)^2}\right],\label{eq:mdn}
    \end{align}
    where we consider the logarithm of the mass for numerical stability and the insight provided by the Press-Schechter formalism on the mass function.
    For convenience, we will denote the parameters of the $i^{\textrm{th}}$ distribution as $\alpha(\boldsymbol{\psi}, \boldsymbol{\theta}_i^{\boldsymbol{\alpha}})\equiv \alpha_i$, $\mu(\boldsymbol{\psi}, \boldsymbol{\theta}_i^{\boldsymbol{\mu}})\equiv \mu_i$ and $\sigma(\boldsymbol{\psi}, \boldsymbol{\theta}_i^{\boldsymbol{\sigma}})\equiv \sigma_i$.
    The parameters of our mixture of Gaussians depend on $\boldsymbol{\psi}$ and the trainable parameters $\boldsymbol{\theta}^\textrm{\scriptsize MDN} = \left\{\boldsymbol{\theta}^{\boldsymbol{\alpha}}, \boldsymbol{\theta}^{\boldsymbol{\mu}}, \boldsymbol{\theta}^{\boldsymbol{\sigma}}\right\}$, where each $\boldsymbol{\theta}^{\boldsymbol{\alpha}, \boldsymbol{\mu}, \boldsymbol{\sigma}}=\left\{w_i^{\alpha, \mu, \sigma}, b_i^{\alpha, \mu, \sigma}\big|~i\in [0, N)\right\}$.
    The parameters of the Gaussians are calculated as such
    \begin{align}
        \alpha_i & = \text{softplus}(w_{i}^\alpha\boldsymbol{\psi}+b^\alpha_i),\\
        \mu_i & = \left\{\begin{array}{lll}
            w_{i}^\mu\boldsymbol{\psi}+b^\mu_i&\textrm{if}&i=0\\
            \textrm{Max}\left[0,~ w_{i}^\mu\boldsymbol{\psi}+b^\mu_i\right]+\mu_{i-1}&\textrm{if}&i>0\label{eq:mu}\\
            \end{array}\right.,\\
        \sigma_i & = \text{softplus}(w_{i}^\sigma\boldsymbol{\psi}+b^\sigma_i)\;,
    \end{align}
    where the function $\text{softplus}$ is defined as $\text{softplus}(x) \equiv \log\left(1+\exp x\right)$.
    The softplus function is used to ensure the positivity of both the amplitude and standard deviations of the mixture density network, while being differentiable at any point.
    The means of the MDN are ordered by amplitude from smallest to largest.
    This is required to remove the degeneracy introduced when removing the softmax constraint on the amplitudes.
    
    \medskip
    Since we have some prior on the function, we decide to use a shifted coordinate system for the weights, making use of our knowledge of the halo mass distribution function (seen in Figure~\ref{fig:hmdf}).
    We therefore include an initial amplitude $\alpha^\textrm{init}$, mass threshold $M_\textrm{th}$, and initial width $\sigma^\textrm{init}$ such that
    \begin{align}
        \boldsymbol{b}^\alpha & \to \boldsymbol{b}^\alpha +\alpha^\textrm{init}\\
        b_0^\mu & \to b_0^\mu + \log\left(M_\textrm{th}\right)\\
        \boldsymbol{b}^\sigma & \to \boldsymbol{b}^\alpha + \sigma^\textrm{init} 
    \end{align}
    which allows all the parameters of the MDN, $\boldsymbol{\theta}^\textrm{\scriptsize MDN}$, to be approximately centred on zero.
    This will be useful when sampling the parameters of the neural bias model since the prior is simple to implement (discussed further in Section~\ref{ssec:HMCLET}).
    
    \medskip
    Using equation~\eqref{eq:mdn} as the halo mass distribution function we can rewrite the log-likelihood (in equation~\eqref{eq:likelihood}) as
    \begin{multline}
        \mathcal{L}  =\hskip-1.2em \sum_{j\in\textrm{catalogue}}\hskip-1em\log\left[\sum_i^N\frac{\alpha_{i,j}}{\sqrt{2\pi\sigma_{i,j}^2}}\exp\left[-\frac{\left(\log(M_j) - \mu_{i,j}\right)^2}{2\sigma_{i,j}^2}\right]\right]\\
         - V\hskip-1.2em\sum_{j\in\textrm{voxels},i=1}^N\hskip-0.5em\frac{\alpha_{i,j}}{2}\exp\left[\frac{\sigma_{i,j}^2}{2}\right]\textrm{erfc}\left[\frac{\log\left(M_\textrm{th}\right) - \mu_{i,j} - \sigma_{i,j}^2}{\sqrt{2\sigma_{i,j}^2}}\right].\label{eq:Ls}
    \end{multline}
    This is the effective likelihood surface that we wish to explore to be able to infer the distribution of parameters in the neural bias model at the same time as the initial phases of the dark matter distribution.
    
    \medskip
    Removing the constraint on the amplitudes of the distributions means that the number of distributions can no longer self-regulate.
    Therefore, we must choose a fixed number of distributions, determined by the expected shape of the function.
    It is possible to make use Bayesian optimisation and model comparison to perform the regularisation, and this will be studied further in future works.
    
    \subsection{HMCLET}
    \label{ssec:HMCLET}
    
    To obtain a Bayesian interpretation of the neural bias model (and the initial phases of the density field) we need to infer the distribution of the parameters of the model.
    The landscape of the likelihood surface (in equation~\eqref{eq:Ls}) is extremely flat in the directions of the parameters of the neural bias model and are highly correlated, often in unpredictable ways.
    As such we need to use specific techniques to be able to effectively traverse the likelihood surface.
    We use Hamiltonian Monte Carlo~\citep[HMC, also known as hybrid Monte Carlo][]{Duane:1987} to draw samples from the target conditional posterior distribution of the parameters, given the matter density field.
    HMC is a Markov chain Monte Carlo method where the proposed states are dictated by a \emph{momentum}, $\boldsymbol{p}$, i.e. the first order gradient information of the target distribution, and the acceptance rate is kept high via conservation of energy momentum.
    In particular, the Hamiltonian is defined as the negative log-probability of the distribution, $\mathcal{P}(\boldsymbol{\theta},\boldsymbol{p})$, of model parameters, $\boldsymbol{\theta}$ and momenta, $\boldsymbol{p}$ 
    \begin{align}
        \mathcal{H}(\mvec{\theta}, \mvec{p}) &= \mathcal{P}(\mvec{\theta}, \mvec{p})\nonumber\\
        &=\mathcal{V}(\mvec{\theta}) + \mathcal{K}(\mvec{p})\nonumber\\
        &=\mathcal{L}(\mvec{\theta}|\mvec{d}) - \log\left[\pi(\mvec{\theta})\right] + \frac{1}{2}\mvec{p}^T\mmat{M}^{-1}\mvec{p}+\textrm{constant}.
    \end{align}
    $\mathcal{K}(\boldsymbol{p})$ is a \emph{kinetic energy} with a mass matrix, $\mmat{M}$, describing the correlation between parameters.
    $\mathcal{V}(\mvec{\theta})$ is a potential energy formed from the negative logarithm of the likelihood (in equation~\eqref{eq:Ls}) and the prior, $\pi(\mvec{\theta})$ on the parameters.
    The state $\boldsymbol{z}=\{\boldsymbol{\theta}, \boldsymbol{p}\}$ is found by solving the ordinary differential equation (ODE) derived from Hamiltonian dynamics
    \begin{align}
        \dot{\mvec{\theta}} & = \mmat{M}^{-1}\mvec{p} \label{eq:H1}\\
        \dot{\mvec{p}} & = -\nabla\mathcal{V}(\mvec{\theta}),\label{eq:H2}
    \end{align}
    where the dots are derivatives in time (as introduced for the momenta).
    Proposals of the $i^\textrm{th}$ parameters, $\boldsymbol{\theta}_i$, are generated by drawing a momentum from a proposal distribution, $\boldsymbol{p}_i\leftarrow\mathcal{N}(\boldsymbol{0}, \boldsymbol{M})$, and evolving these using equations~\eqref{eq:H1} and~\eqref{eq:H2} to obtain $\boldsymbol{z}^*=\{\boldsymbol{\theta}^*, \boldsymbol{p}^*\}$.
    The acceptance condition for the Metropolis-Hasting procedure is obtained by computing the difference in energies between the $i^\textrm{th}$ state and the proposed state
    \begin{equation}
        \textrm{Acceptance probability} = \textrm{Min}\left[\exp(\Delta\mathcal{H}), 1\right],
    \end{equation}
    where $\Delta\mathcal{H}=\mathcal{H}(\boldsymbol{\theta}_i, \boldsymbol{p}_i)-\mathcal{H}(\boldsymbol{\theta}^*, \boldsymbol{p}^*)$ arises from the discretisation of solving Hamilton's equations.
    If the equations were solved exactly (the Hamiltonian is conserved), then every single proposal is accepted.
    It is typical to use $\epsilon$-discretisation (leapfrog method) to solve the ODE, where $\epsilon$ describes the step size of the integrator~\citep{Verlet:1967}.
    Smaller step sizes result in higher acceptance rate at the expense of longer computational times of the integrator, whilst larger step sizes result in shorter integration times, but lower acceptance.
    
    \medskip 
    Because the likelihood described in equation~\eqref{eq:Ls} is flat with sharp edges, the momentum of the model parameters can cause the integrator to step away from the domain of existence of the gradients of the likelihood.
    To prevent this, we choose a Gaussian prior on each of the parameters $\boldsymbol{\theta}=\{\boldsymbol{\theta}^\textrm{\scriptsize NPE}, \boldsymbol{\theta}^\textrm{\scriptsize MDN}\}$, centred on zero with identical widths
    \begin{equation}
        \pi(\boldsymbol{\theta}) \propto \exp\left[-\frac{|\boldsymbol{\theta}|^2}{2\sigma^2_\textrm{prior}}\right].
    \end{equation}
    We could choose more general priors independently for each parameter, but this is difficult since we do not have prior information on the scale of the weights of the network. 
    We are able to set the mean of the Gaussian prior to zero by ensuring that all trainable parameters of the neural bias model are close to zero via the use of the initial amplitude, width and mass threshold mentioned in section~\ref{ssec:MDN}.
    As such, we just have to choose the Gaussian prior to be wide enough to allow plenty of freedom for parameter value exploration whilst preventing the parameters of the neural bias model from becoming extremely large. 
    This also implies that the total probability distribution is now ensured to be proper.
    
    \medskip
    We also use an adaptation to the usual HMC paradigm, the quasi-Newtonian HMC~\citep[QNHMC][]{Fu:2016}.
    This is because there is very little \emph{a priori} knowledge about the correlations between the parameters of the neural bias model, especially those in the neural physical engine, and therefore the ODE is exceptionally stiff.
    With the QNHMC we make use of the second order geometric information of the target distribution as well as the gradient.
    This additional information can be efficiently approximated using quasi-Newtonian methods.
    The QNHMC modifies equation~\eqref{eq:H1} and \eqref{eq:H2} to
    \begin{align}
        \dot{\boldsymbol{\theta}} &=\mmat{B}\mmat{M}^{-1} \boldsymbol{p} \\
        \dot{\boldsymbol{p}} &=-\mmat{B}\nabla\mathcal{V}(\boldsymbol{\theta})
    \end{align}
    where $\boldsymbol{B}$ is an approximation to the inverse Hessian derived from the L-BFGS technique \citep{Liu89onthe} found using quasi-Newton methods
    \begin{equation}
        \mmat{B}^*=\left(\mathbb{I}-\frac{\boldsymbol{s}_i\boldsymbol{y}_i^T}{\boldsymbol{y}_i^T\boldsymbol{s}_i}\right)\mmat{B}_i\left(\mathbb{I}-\frac{\boldsymbol{y}_i\boldsymbol{s}_i^T}{\boldsymbol{y}_i^T\boldsymbol{s}_i}\right)+\frac{\boldsymbol{s}_i\boldsymbol{s}_i^T}{\boldsymbol{s}_i^T\boldsymbol{y}_i},
    \end{equation}
    where $\boldsymbol{s}_i=\boldsymbol{\theta}^*-\boldsymbol{\theta}_i$, $\boldsymbol{y}_i=\nabla\mathcal{V}(\boldsymbol{\theta}^*)-\nabla\mathcal{V}(\boldsymbol{\theta}_i)$ and $\mathbb{I}$ is the identity matrix.\footnote{For large dimensions, more computationally and memory efficient methods can be used~\citep{Nocedal:2006}.}
    The inverse Hessian effectively rescales the momenta and parameters such that each dimension has a similar scale and thus the movement around the likelihood surface is more efficient and the produced proposals are less correlated. 
    The mass matrix is still present to set the dynamical time scales of the ODE problem along each direction. The rationale behind the choice of the mass matrix is indicated in Appendix~\ref{app:mass_qnhmc}.
    Note that the approximate inverse Hessian varies with proposal, but is kept constant whilst solving the ODE.
    Obtaining $\mmat{B}^*$, is extremely efficient because both $\boldsymbol{s}_i$ and $\boldsymbol{y}_i$ are calculated when solving equations~\eqref{eq:H1} and~\eqref{eq:H2} using leapfrog methods.
    We start with an initial inverse Hessian $\mmat{B}_0=\mathbb{I}$ and allow it to adapt to the geometry of the space.
    Although this requires an estimate of the mass matrix initially, the rescaling of the momenta via $\mmat{B}$ allows us to be fairly ambiguous about its value.
    In essence, this all occurs during the burn-in phase of the HMC sampler.

\section{BORG framework}
    \label{sec:BORG}  
        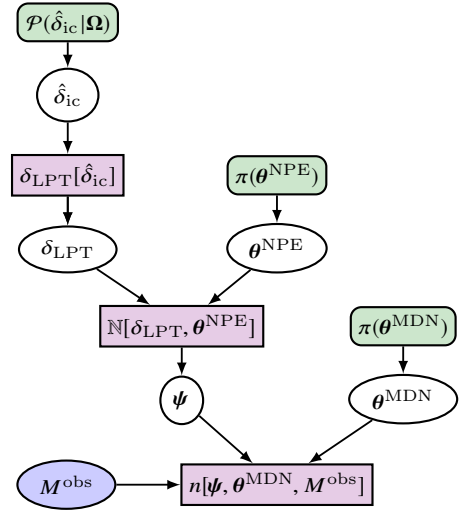
\begin{figure}
        \begin{center}
            \input{diagram_tikz.tex}
        \end{center}
        \caption{Schematic of the \borg{} algorithm with the neural bias model.
        Initial conditions for the density field in Fourier space, $\hat{\delta}_\textrm{ic}$, are drawn from a prior given a cosmology $\boldsymbol{\Omega}$, $\mathcal{P}(\hat{\delta}_\textrm{ic}| \boldsymbol{\Omega})$.
        These are then evolved forward using a deterministic prescription, in this example using LPT.
        The evolved field is then transformed further using the neural physical engine $\mathbb{N}$ which requires parameters $\boldsymbol{\theta}^\textrm{NPE}$ which are drawn from a prior $\pi(\boldsymbol{\theta}^\textrm{NPE})$.
        This provides a field $\boldsymbol{\psi}$ from which the halo mass distribution function can be described using the mixture density network with parameters $\boldsymbol{\theta}^\textrm{MDN}$ drawn from a prior $\pi(\boldsymbol{\theta}^\textrm{MDN})$.
        This halo mass distribution function is then compared to the masses of halos $M^\textrm{obs}$ from the observed halo catalogue.} 
        \label{fig:mdn_diagram}
    \end{figure}
    Our ability to sample the parameters of the neural bias model builds upon the previously developed \borg{} algorithm.
    \borg{} aims to the analyse the 3D cosmic matter distribution at linear and non-linear scales of structure formation from galaxy surveys \citep[see e.g.][]{Jasche:2013, Jasche:2015, Lavaux:2016}.
    Explicitly, the \borg{} algorithm fits 3D models of gravitational structure formation to data.
    Via the introduction of a physical model of gravitational structure growth, the inference of the non-linear dark matter distribution today can be posed as a statistical initial conditions problem. 
    To do so, the \borg{} algorithm seeks to infer the cosmic initial conditions from which present 3D structures in the distribution of galaxies have formed via non-linear gravitational mass aggregation.
    
    \medskip
    The \borg{} algorithm explores the posterior distribution of large scale structures, consisting of a Gaussian prior for the initial density field at an initial scale factor of $a=10^{-3}$ and a choice of bias model and likelihood metric at scale factor a = 1.
    The evolution of the initial density fields can be related to the present galaxy (or halo) distribution via a first or second order Lagrangian perturbation theory (LPT or 2LPT) or a full particle mesh model of gravitational structure formation \citep[for details see][]{Jasche:2013,Jasche:2019}. 
    By exploiting non-linear structure growth models the \borg{} algorithm naturally accounts for the filamentary structure of the cosmic web typically associated with higher-order statistics induced by non-linear gravitational processes. 
    Furthermore, the posterior distribution accounts for systematic and stochastic uncertainties, such as survey geometries, selection effects, unknown noise and galaxy biases as well as foreground contamination \citep[see e.g.][]{Jasche:2013, Jasche:2015, Lavaux:2016, Jasche:2017}.
    
    \medskip
    In this work we use the \borg{} algorithm to sample the initial conditions of the dark matter density field using LPT to evolve the field to the dark matter conditions today and then rely on the neural bias model to sample the parameters of the neural physical engine to provide a field which is most informative about the halo mass distribution function inferred from a halo catalogue.
    A detailed schematic of the interconnection between parts of the model is shown in Figure~\ref{fig:mdn_diagram}.
    The stochastic uncertainties are assumed to be Poissonian using the likelihood in Equation~\eqref{eq:Ls}, and further study into direct learning of the deviations from Poissonity will be studied in an upcoming work. 
    
\section{Results}
    \label{sec:results}
    
    To examine the techniques developed in this paper we will consider a relatively simple mock run using a simulated dark matter distribution and minimal working neural bias model.
    
    \subsection {VELMASS simulation} 
    The halo catalogue which we use in this work comes from the \textsc{velmass} suite.
    It is comprised of 10 cosmological simulations, 9 of which are probing slightly different variations of a selection of cosmological parameters whilst using the same initial phases \citep[for full details of the suite of simulations see][]{Ramanah:2019}. 
    We recall here the salient features which are relevant for this work.
    The simulation that we use in this work assumes a Planck-like cosmology~\citep{Planck:2015} with $\Omega_\text{m}=0.315$, $\Omega_\text{b}=0.049$, $H_0=68\,\mathrm{km~s}^{-1}\mathrm{ Mpc}^{-1}$, $\sigma_8=0.81$, $n_{\mathrm{s}}=0.97$ and $Y_\mathrm{He}=0.248$ (named ``central'' or $\Omega$ simulation).
    The power spectrum is obtained through the analytic prescription of \citet{Eisenstein:1999}, and the initial conditions are generated by \textsc{music}~\citep{Hahn:2011}. 
    
    \medskip
    The cosmological simulation covers a volume of 2000$h^{-1}$~Mpc with 2048$^3$ particles tracing dark matter, initialised at a redshift $z=50$ and evolved to present time with \textsc{gadget2}~\citep{Springel:2005}, adopting a softening length for gravity equal to 48$h^{-1}$~kpc corresponding to $1/20^\textrm{th}$ of the mean interparticle separation.
    The \textsc{rockstar} halo finder algorithm \citep{Behroozi:2013} was subsequently employed to extract the halos from the simulation and generate the halo catalogue. 
    The particle mass resolution is 8.10\Msun.
    
    \medskip
    Using a patch of the central \textsc{velmass} simulation of side $250h^{-1}$Mpc gridded onto a $64^3$ grid,\footnote{High resolution results over the whole simulation will be reserved for future papers studying optimal forms of the neural bias model and studies into direct likelihood estimation. $64^3$ is more than sufficient for indicating the methodology of the techniques presented in this paper.} and a fairly sampled set of halos from the corresponding halo catalogue, we attempt to constrain both the initial phases of the dark matter density field as well as the parameters of the neural bias model.
    
    \subsection{Neural bias model}
    
    In this work we focus on a minimal model using a single $\ell=0$ kernel with extent of $3^3$ voxels ($\sim12h^{-1}$Mpc/side).
    This is sufficient for studying the first order effect of the beyond-local density environment.
    We also use a single softplus activation function on the output feature map from the $\ell=0$ kernel which provides us with the non-linearity necessary to infer the parameters of the MDN.
    Further studies into the optimal architecture for extracting all the relevant information from the density field about the halo mass distribution function will be considered in future works. 
    Thanks to the simplicity of the neural physical engine considered in this paper, we only introduce 5 trainable parameters, $\boldsymbol{\theta}^\textrm{\scriptsize NPE}$.
    We centre the parameters of the MDN, $\boldsymbol{\theta}^{\textrm{\scriptsize MDN}}$, on zero by using an initial amplitude $\alpha^\textrm{init}=\log(1\times10^{-3})$, mass threshold $M_\textrm{th}=2\times10^{12}$\Msun, and initial width $\sigma^\textrm{init}=\log(1\times10^{3})$.
    Due to the number of visible features in the halo mass distribution function (in figure~\ref{fig:hmdf}) we determine that two Gaussians are sufficient to model the halo mass distribution function.
    As such, the total number of parameters for the MDN is 12, so the neural bias model has 17 parameters to infer.
    Further study into the number and types of distributions optimal for extracting all of the information from the density field about the halo mass distribution function will be left to future work. 
    We chose a prior width of $\sigma_\textrm{prior}=10$ to allow for a range of possible parameter values in the neural bias model whilst preventing numerical stability issues if they become extremely large.
    A posteriori verification that the prior has no practical impact on the inference is shown in Figure~\ref{fig:weights}, since the amplitude of the weights are all well within the 1$\sigma$ region of the Gaussian prior.
    
    \medskip
    The neural bias model is written using the \textsc{julia} \citep{bezanson_julia:2017} interface to \textsc{TensorFlow}~\citep{Abadi:2016, Malmaud:2018}.
    This is embedded into the \textsc{hmclet} which is one sub-block of the \borg{} algorithm.
    A skeleton of the \textsc{julia} code is available along with this paper.\footnote{\url{https://github.com/tomcharnock/neural_bias_model}}
    
    \medskip
    Since running the HMC is computationally expensive, especially when sampling the initial density field, we pretrain the neural bias model using stochastic gradient descent to find pseudo-maximum likelihood estimates of the weights.
    These parameters are then used in the initialisation of the neural bias model in the \textsc{hmclet} to help prevent a long burn-in.
    This option is available in this case since we have both the true density field and halo catalogue generated from that field.
    However, the pretraining would not be possible when constraining the network on real data, where the density field is not available.
    In this case the burn-in of the \textsc{hmclet} would be much longer and more computationally demanding.

    \subsection{Assessment of the model}
    \begin{figure} 
        \centering
        \includegraphics{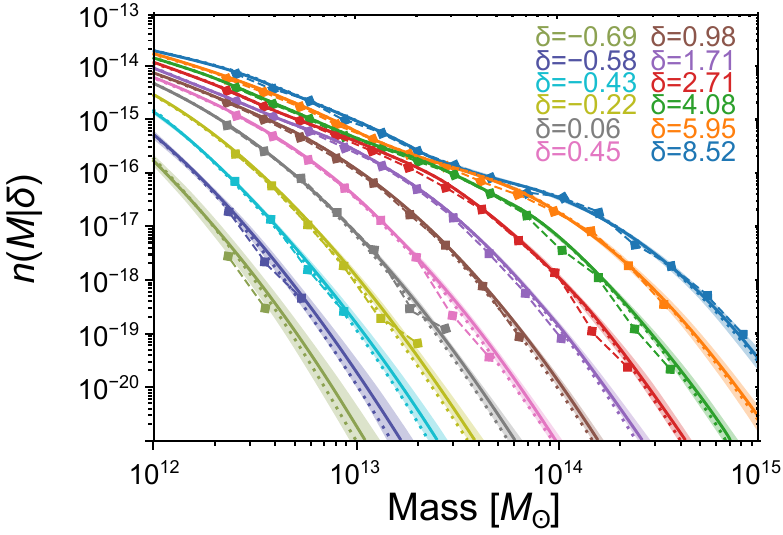}
        \caption{The halo mass distribution function as a function of mass (in solar masses).
        The diamonds connected by a dashed line indicates the number density of halos from the central \textsc{velmass} halo catalogue of a given mass, where the different colours represent the value of the density environment for those halos.
        The lines higher in number density correspond to the more dense regions, i.e. there are more large halos in denser environments.
        The solid lines show the mean halo number density from samples (taken from the Markov chain) from the neural bias model, with the shaded bands as the 1$\sigma$ deviations of these samples.
        The dotted line indicates the number density of halos of a given mass using the initialisation values of the parameters from the neural bias model.
        There is a subtlety in the density environments.
        The diamonds indicate the number density of halos in a voxel whose density is equal to the numbers shown in the legend, whilst the neural bias model takes a $3^3$ patch whose average density is equal to the numbers shown in the legend.
        We see that there is a very good agreement between the observed halo number density and that obtained by the neural bias model.
        We can also see that the sampling captures the distribution of possible number densities about the observed data.
        Furthermore, the fact that the shape of the function changes as a response to the change in density environment is an indication that the non-linearity of the tracer bias is captured by the neural bias model.}
        \label{fig:hmdf}
    \end{figure}
        \begin{figure*}
        \centering
        \includegraphics[width=\textwidth]{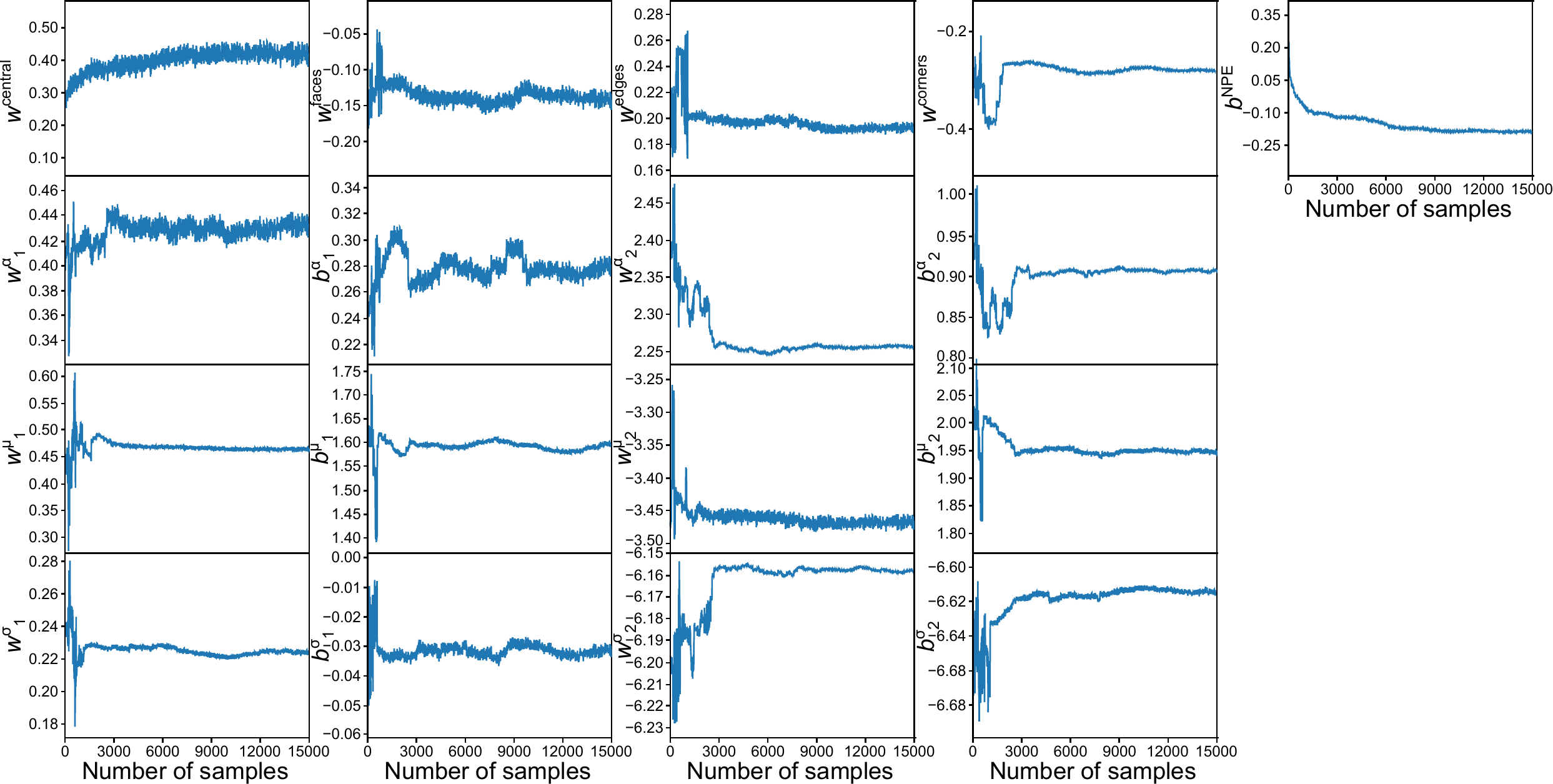}
        \caption{The Markov chain of sampled parameters of the neural bias model.
        The top row shows, from left to right, the values of the sampled weights of the central, face, edge and corner of the $\ell=0$ convolution kernel respectively.
        The final subplot on the top row shows the overall bias parameter of the neural physical engine.
        The lower three rows show the sampled parameter values of the MDN.
        The second row shows the weight and bias which parameterise the amplitude of the first and then the second distribution of the mixture of Gaussians.
        Likewise the third and fourth rows show the weights and biases parameterising the mean and standard deviation of the first and second distributions respectively.
        We can see that the chain wanders quite wildly until around $\sim3000$ samples, at which point the chain is properly burnt-in and the samples are really being drawn from the posterior distribution.}
        \label{fig:weights}
    \end{figure*}
    First, we consider the sampling of the parameters of the neural bias model.
    The obtained halo mass distribution function is shown in Figure~\ref{fig:hmdf}, and the values of the sampled parameters of the model as a function of the Markov chain is shown in Figure~\ref{fig:weights}.
    
    \medskip
    In Figure~\ref{fig:hmdf}, the observed number density of halos from the central \textsc{velmass} simulation halo catalogue is plotted using diamonds (with dashed lines connecting them to help visualisation).
    From top to bottom in number density, the different colours represent decreasing density environments, i.e. voxels with higher densities have a larger halo number density in those voxels.
    We analyse the effectiveness of the neural bias model by providing it with density patches (of size $3^3$ voxels) with an average density equivalent to the values in individual voxels to compare to the halo mass distribution function from the halo catalogue.
    These $3^3$ volumes are randomly drawn to provide a variety of different local patches.
    The mean result of the neural bias model from the Markov chain is shown using a solid line, and the filled area represents the 1$\sigma$ deviations from the samples of parameter values of the model.
    For completeness, we also show the initial values of the parameters of the neural bias model, obtained using stochastic gradient descent, via the dotted line.
    We can see that there is extremely good agreement between the observed halo number density and that obtained using the neural bias model.
    In particular, we notice that the two components of the MDN fit the complexities of the halo mass distribution function well.
    The sampling is also well contained about the observed data, and improved over the initial neural bias model parameter values.
    The fact that the response to different density environments can be seen provides us with confirmation that the non-linearity of the tracer bias is built into our neural bias model.
    
    \medskip
    Turning our attention to Figure~\ref{fig:weights}, we can see the sampling of the parameters of the neural bias model.
    The top row shows the free parameters of the neural physical engine, with the first four subplots showing the weights of the central, face, edge and corner of the $\ell=0$ kernel respectively.
    The fifth subplot is the overall bias to the neural physical engine which sets the response scale for activated feature map which will be passed to the mixture of Gaussians.
    The second, third and fourth rows show the weight and bias values for the amplitude, mean and the standard deviation of the the mixture of Gaussians respectively.
    The first two columns show the weight and bias for the first Gaussian of the mixture and the third and fourth columns show the weight and bias for the second Gaussian.
    Overall we can see that the values of the samples vary a lot during the first $\sim 3000$ samples, after which burn-in ends and the \textsc{hmclet} starts to truly sample well from the posterior.
    The variation in the samples during burn-in occurs even though we pre-initialise the parameters of the neural bias model using stochastic gradient descent.
    This happens because the rescaling of the momenta in the \textsc{hmclet} is being learned via the QNHMC, and because the initial density field is not yet conditioned on the observed halo catalogue.
    Interestingly, we also see that there is some change between the initial and average parameter values.
    This could be caused by the stochastic gradient descent not properly finding the minimum of the log-likelihood, but which is achieved by the \textsc{hmclet}.
    It is more likely to be due to the fact that the density field used to constrain the parameters using stochastic gradient descent is a full N-body simulation, whilst only the LPT field is used in the density sampler, showing that the neural bias model is able to adapt to the missing information due to using only the approximate evolution.
    We can see that our choice of prior width on the parameter values has not affected the posterior since none of the parameters have vanished, and in fact can be relatively large (in the case of the weights and biases for the mean and standard deviation of the second distribution).

    \medskip
    \begin{figure}
        \centering
        \includegraphics[width=8cm]{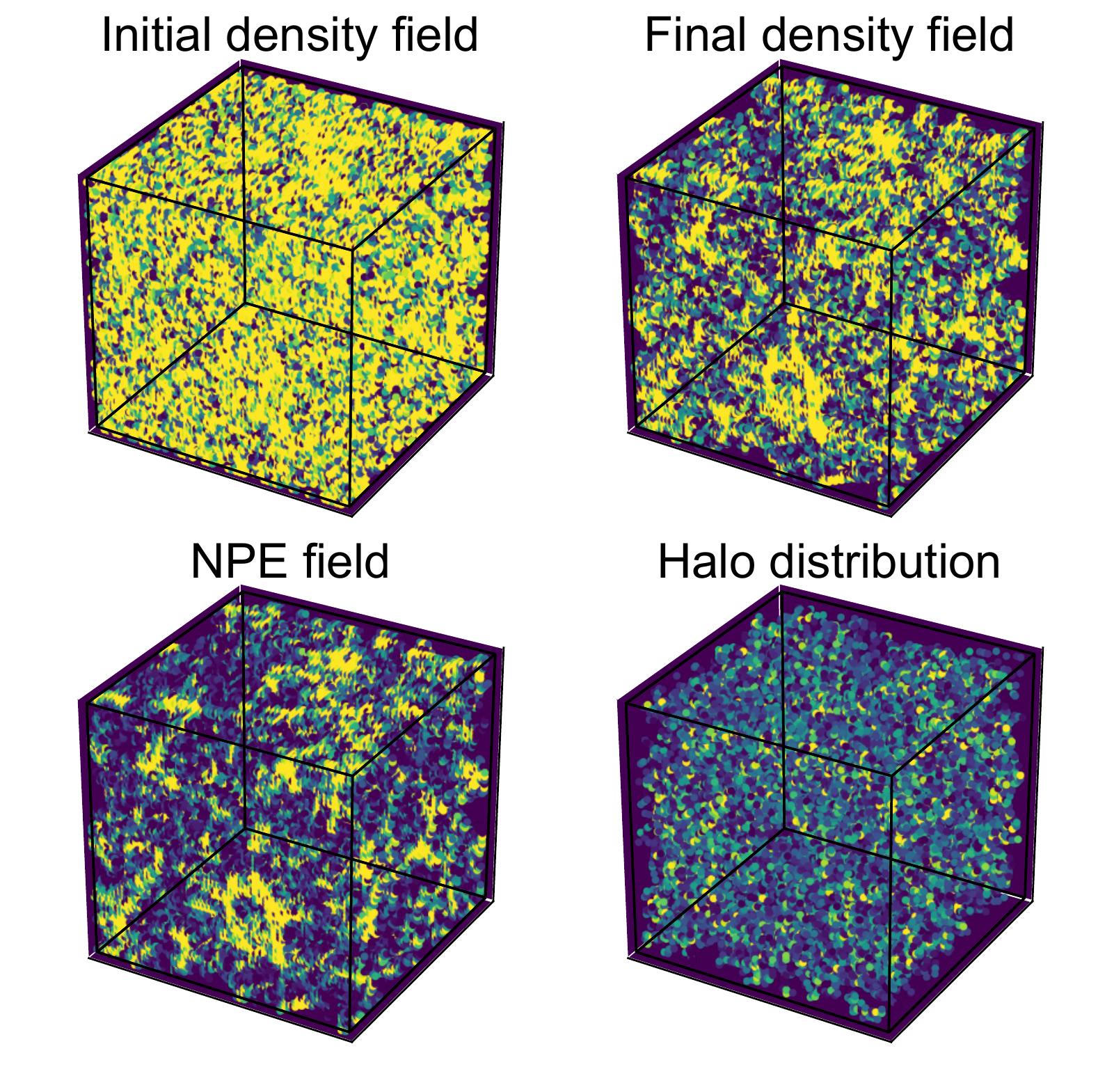}
        \caption{3D projections of relevant fields.
        The upper left box shows the 3D projection of the initial dark matter distribution.
        The upper right box shows the same projection of the dark matter distribution evolved to a scale factor $a=1$ using LPT.
        The lower left box shows the output of the neural physical engine and, for completeness, we show the logarithm of the mass distribution of the halo catalogue in the lower right box.
        The three density boxes all use the same colour scale.
        We can see that the production of the non-linear features of evolution by comparing the initial and final density fields (in the top row), whilst we see the enhancing effect of the non-linear structures due to the neural physical engine in the lower left box.
        Note the stochastic nature of the halo distribution obtained from the observed halo catalogue compared to the field obtained from the neural physical engine.}
        \label{fig:3D}
    \end{figure}
    By looking at the average samples in the first four subplots on the top row of Figure~\ref{fig:weights} we can introspect our physical neural engine.
    In this case, since we only have one kernel, it is very easy to see the effect of the kernel on the density field.
    This type of kernel resembles that of a contrast-increasing kernel, which enhances dense regions and washes out underdense areas.
    In fact, if we look at Figure~\ref{fig:3D} we can see, qualitatively, the effect of the neural physical engine.
    The upper left box shows the inferred initial conditions of the dark matter density field, which is then evolved forward using LPT to obtain the box on the upper right.
    The lower left box then shows the output of the neural physical engine.
    The lower right box shows the distribution of mass from the halo catalogue, for completeness.
    We can see the aforementioned increase in contrast of the density field which the neural physical engine provides.
    The existence of non-zero value for the faces, edges, corners and central part of the $\ell=0$ kernel shows us that beyond-local information from the density field is important for the fitting of the halo mass distribution function.
    The neural bias model therefore makes use of information from the surrounding regions of each voxel to improve the quantification of the number density of halos with a given mass in each voxel.
    
    \medskip
    The halo distribution should be a stochastic sampling of the inferred halo mass distribution function provided by the neural bias model.
    In the current design, this sampling would be described via a Poisson distribution.
    Deviations between such a realisation and the halo distribution from the catalogue could be due the nature of the true likelihood which is expected to be non-Poissonian.
    
    \medskip
    \begin{figure}
        \centering
        \includegraphics{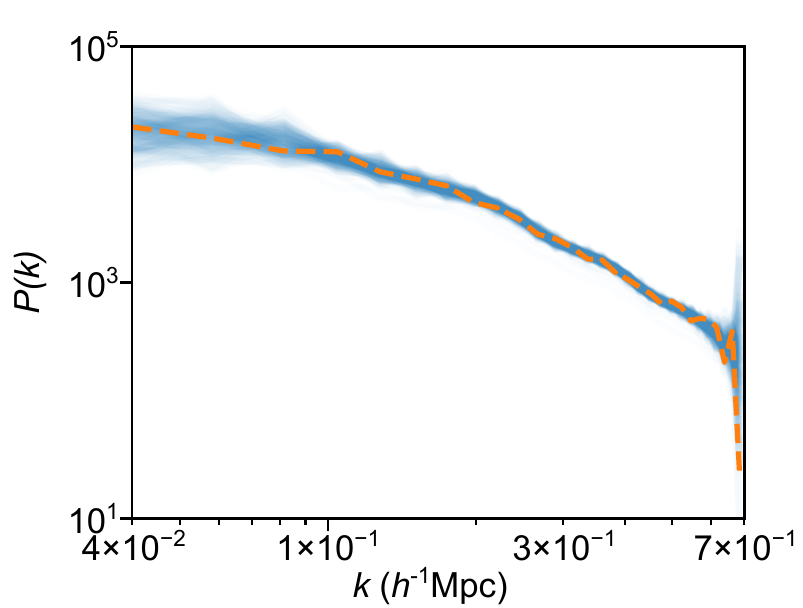}
        \caption{The power spectrum of initial conditions of the dark matter density field.
        The blue transparent lines show the power spectrum from the posterior samples.
        The orange dashed line shows the power spectrum of the prior initial dark matter distribution.
        We can see that the inferred power spectra are consistent with the prior initial power spectrum without any spurious features.}
        \label{fig:Pk}
    \end{figure}
    Via the \borg{} algorithm, we also sampled the initial conditions of the dark matter density field (seen in the upper left box of Figure~\ref{fig:3D}).
    We show in Figure~\ref{fig:Pk} the power spectrum of the initial conditions.
    The orange dashed line is the prior power spectrum from which the initial density field is drawn.
    The blue transparent lines show the initial power spectrum of the dark matter density field from the posterior samples inferred via \borg{}.
    We can see that the inferred power spectra are consistent to the prior, and there are no spurious features.
    This indicates that the sampling of the density field is self-consistent with the prior provided and the data used to generate the halo catalogue.
    
    \medskip
    We compare the quality of the neural bias model, $\mathcal{M}_\textrm{Neural bias}$, to the bias model of \citet{Neyrinck:2014}, $\mathcal{M}_\textrm{Neyrinck}$, where the Poisson rate for the realisation of the halo field is given by
    \begin{equation}
        \lambda = \bar{N}(1+\delta_\textrm{LPT})^\alpha\exp\left[-A(1+\delta_\textrm{LPT})^{-\varepsilon}\right].
    \end{equation}
    This model is superior to a simple linear bias model since it enhances the contrast between dense regions and voids.
    There are four free parameters, $\boldsymbol{\theta}_\textrm{Neyrinck}=\{\bar{N}, \alpha, A, \varepsilon\}$.
    By evaluating the value of the normalised likelihood given the \textsc{velmass} halo catalogue and a single realisation of the $\delta_\textrm{LPT}$ field at the maximum likelihood values of the parameters, $\hat{\boldsymbol{\theta}}_\textrm{Neyrinck}$, and comparing it to the normalised likelihood for the neural bias model at the maximum likelihood values of its parameters, $\hat{\boldsymbol{\theta}}_\textrm{Neural bias}$, we can get a sense of how much better the neural bias model performs.
    We find that the relative likelihood ratio~\citep{Neyman:1933} is \begin{align}
        B(\mathcal{M}_\textrm{Neural bias}, \mathcal{M}_\textrm{Neyrinck}) & = \frac{\mathcal{L}(d|\hat{\boldsymbol{\theta}}_\textrm{Neural bias}, \mathcal{M}_\textrm{Neural bias})}{\mathcal{L}(d|\hat{\boldsymbol{\theta}}_\textrm{Neyrinck}, \mathcal{M}_\textrm{Neyrinck})}\nonumber\\
        & \approx \exp[2000].
    \end{align}
    which is decisive.

\section{Conclusions}
    \label{sec:conclusions}
    
    We have presented a neural bias model: a physically motivated neural network-like architecture which maps dark matter density fields to the halo mass distribution function.
    In doing so we have used a swathe of new techniques including novel architectures, such as the multipole expansion in convolutional kernels, and well adapted sampling methods (such as the QNHMC) to provided a Bayesian interpretation of the parameters of the neural bias model.
    
    \medskip
    Most importantly, we have shown how physical principles allow us to build extremely efficient neural networks whose parameters can be sampled to provide a truly Bayesian network.
    The neural bias model becomes part of the forward physical model meaning that the posterior is only conditional on the architecture in the same way as the model that describes the data.
    No training data is necessary since the weights are inferred directly from the observed data.
    
    \medskip
    We have found that an exceptionally simple neural bias model constructed using a single, rotationally symmetric convolutional kernel and a mixture density network consisting of a mixture of two Gaussians can effectively model the halo mass distribution function.
    This neural bias model contains the non-linear response of different density environments and is able to make use of the information from neighbouring patches of the universe to better predict the abundance of halos of given masses.
    Furthermore, the parameters of the model have been inferred using only forward simulations of the dark matter density field and the observed halo catalogue, providing us with a completely Bayesian interpretable network.
    The simplicity of the model has also allowed us to introspect the neural bias model to see that the kernel enhances dense regions since this informs the halo mass distribution function about where halos are more likely to be abundant. 
        
    \medskip
    This work is proposed as the initial work in a suite of follow-up studies including using Bayesian optimisation routines to find the optimal architecture for the neural physical engine and the number and types of distribution for the mixture density network which allow us to extract maximal information about the halo mass distribution function from the density field.
    We shall also study the effects of the likelihood that we use to evaluate the density field given the halo catalogue via the replacement of the halo mass distribution function with a neural density estimator emulating the unknown likelihood.
    This will allow us to search beyond-Poissonity likelihoods.
    The culmination of this suite of works will be to analyse real cosmological survey data using \borg{} with a completely agnostic neural bias model, which can be marginalised out to provide constraints on the initial distribution of dark matter, coupled with cosmological parameters, independent of the unknown astrophysics which dictates the tracer bias between the dark matter distribution we use and the observable universe.

\section*{Acknowledgements}
    This work was supported by the ANR BIG4 grant ANR-16-CE23-0002 of the French Agence Nationale de la Recherche.
    TC wishes to thank \textsc{nvidia} for the Quadro P6000 used in this work.
    This work has made use of the Horizon Cluster hosted by Institut d'Astrophysique de Paris.
    This work has been done within the activities of the Domaine d'Intérêt Majeur (DIM) ``Astrophysique et Conditions d’Apparition de la Vie" (ACAV), and received financial support from R\'egion \^Ile-de-France. 
    The work was granted  access  to  the  HPC  resources  of  CINES  (Centre  Informatique National de l'Enseignement Sup\'erieur) under the allocation A0020410153 and A0040410153 made by GENCI for the VELMASS simulations
    This work was done as part of the \textsc{aquila} consortium.\footnote{\url{https://www.aquila-consortium.org/}}
    MJH is supported by NSERC (Canada).

\section*{Contribution}
    TC designed the neural bias model, developed the multipole kernels module, ran the fitting algorithm, contributed to the development of the \textsc{hmclet} and wrote the bulk of the content of the paper.
    GL was the main developer of the \textsc{hmclet} and ran the \textsc{velmass} simulation suite.
    BDW first proposed the idea of including a neural network and sampling its parameters as part of the \borg{} algorithm, allowing one to be able to marginalise out uncertainties in the bias model.
    SSB and MJH proposed the idea at the basis of the likelihood for halo distribution.
    JJ proposed the use of the QNHMC algorithm. GL and JJ are the main developers of the \borg{}$^3$ software used in this work.
    TC, BDW and GL are the main proponents of developing neural networks using physical principles.

\section*{Environmental impact}

    This study has made use of 103 single core days and 2 single GPU days on a high performance computing cluster and 90 single core days and 8 full GPU days on a 850W workstation loaded with an \textsc{nvidia} Quadro P6000. 
    This amounts to approximately 1100kWh including cooling and data storage.
    In the Paris metropolitan this would be equivalent to approximately 55kg of CO$_2$.
    
    \medskip
    We have also reused simulations from the \textsc{velmass} suite.
    These simulations were created for the purpose of being a general tool for a wide variety of projects.
    It's longevity reduces its single use cost dramatically.
    This suite of simulations took 23000 single core days at Occigen facility managed by CINES.
    The amounts to approximately 5500kWh and is equivalent to 275kg of CO$_2$ in the H\'erault metropolitan.
    The \textsc{velmass} suite is stored at l'Institut d'Astrophysique de Paris at a cost of around 260kWh/year which is approximately 13kg/year of CO$_2$.
    
    \medskip
    All values have been approximated using the Parliamentary Office of Technology document on Carbon Footprint of Electricity Generation and according to l'Agence Internationale de l'\'Energie.
    Exact figures were not available for the power consumption of the computing facilities, and as such a generous approximation has been considered.
    This is because figures are generally represented in terms of running costs and not in terms of power usage or environmental impact.
    
\bibliographystyle{mnras}
\bibliography{bib}

\appendix
\section{Mass matrix for QNHMC}
\label{app:mass_qnhmc}

The new set of ordinary differential equations introduced by the QNHMC yield substantial modifications to the HMC prescription for the optimal mass matrix to sample the parameter space with low rejection rate.
This can be seen by considering a Gaussian posterior distribution with covariance $\mmat{C}$, then
\begin{equation}
    \mathcal{V}(\mvec{\theta}) = \frac{1}{2}\mvec{\theta}^{T}\mmat{C}^{-1}\mvec{\theta}\;.
\end{equation}
The approximate inverse Hessian of $\mathcal{V}(\mvec{\theta})$, $\mmat{B}$, should satisfy $\mmat{B}\simeq\mmat{C}$.
In this case the two equations \eqref{eq:H1} and~\eqref{eq:H2} becomes
\begin{align}
    \dot{\mvec{\theta}} &= \mmat{C} \mmat{M}^{-1}\mvec{p} \\
    \dot{\mvec{p}} &= -\mmat{C} \mmat{C}^{-1} \mvec{\theta}\nonumber\\& = -\mvec{\theta}\;.
\end{align}
These two equations can be combined to form a single equation
\begin{equation}
    \ddot{\mvec{\theta}} + \mmat{C}\mmat{M}^{-1} \mvec{\theta} = \mvec{0}\;.
\end{equation}
To numerically integrate the above equation with a leapfrog integrator in an optimal way, it is best to choose a mass matrix satisfying 
\begin{equation}
    \mmat{M}=\mmat{C}\;.
\end{equation}  
In most practical cases, we choose a diagonal mass matrix with coefficients that are close to the expected width of the posterior distribution.

\bsp
\label{lastpage}
\end{document}

%% file: diagram_tikz.tex
    \begin{tikzpicture}
    	\pgfdeclarelayer{background}
    	\pgfdeclarelayer{foreground}
    	\pgfsetlayers{background,main,foreground}
    
        \tikzstyle{probability}=[draw, thick, text centered, rounded corners, minimum height=1em, minimum width=1em, fill=darkgreen!20]
    	\tikzstyle{deterministic}=[draw, thick, text centered, minimum height=1.8em, minimum width=1.8em, fill=violet!20]
    	\tikzstyle{variabl}=[ellipse, draw, thick, text centered, minimum height=1em, minimum width=1em]
    	\tikzstyle{data}=[ellipse, draw, thick, text centered, minimum height=1em, minimum width=1em, fill=blue!20]
    
    	\def\blockdist{0.7}
    
        \node (ic) [probability]
        {$\proba(\hat{\delta}_\text{ic}|\mvec{\Omega})$};
        \path (ic.south)+(0,-\blockdist) node (deltaic) [variabl]
        {$\hat{\delta}_\text{ic}$};
        \path (deltaic.south)+(0,-\blockdist) node (deltalpt) [deterministic]
        {$\delta_\text{LPT}[\hat{\delta}_\text{ic}]$};
        \path (deltalpt.west)+(3.5,0) node (PNPE) [probability] {$\pi(\mvec{\theta}^\textrm{NPE})$};
        \path (PNPE.south)+(0,-\blockdist) node (thetaNPE) [variabl] {$\boldsymbol{\theta}^\textrm{NPE}$};
        \path (deltalpt.south)+(0,-\blockdist) node (deltalpt_field) [variabl]
        {$\delta_\text{LPT}$};
        \path (deltalpt_field.south)+(1.5,-\blockdist) node (NPE) [deterministic]
        {$\mathbb{N}[\delta_\text{LPT}, \boldsymbol{\theta}^\textrm{NPE}]$};
        \path (NPE.south)+(0,-\blockdist) node (psi) [variabl]
        {$\boldsymbol{\psi}$};
        \path (NPE.west)+(4,0) node (PMDN) [probability] {$\pi(\mvec{\theta}^\textrm{MDN})$};
        \path (PMDN.south)+(0,-\blockdist) node (thetaMDN) [variabl] {$\boldsymbol{\theta}^\textrm{MDN}$};
        \path (thetaNPE.south)+(0,-\blockdist-\blockdist-\blockdist-\blockdist) node (MDN) [deterministic] {$n[\boldsymbol{\psi}, \boldsymbol{\theta}^\textrm{MDN}, M^\textrm{obs}]$};
        \path (MDN.west)+(-1.5,0) node (M) [data] {$M^\textrm{obs}$};
    
    	\path [draw, line width=0.7pt, arrows={-latex}] (ic) -- (deltaic);
    	\path [draw, line width=0.7pt, arrows={-latex}] (deltaic) -- (deltalpt);
    	\path [draw, line width=0.7pt, arrows={-latex}] (deltalpt) -- (deltalpt_field);
    	\path [draw, line width=0.7pt, arrows={-latex}] (deltalpt_field) -- (NPE);
    	\path [draw, line width=0.7pt, arrows={-latex}] (deltalpt) -- (deltalpt_field);
    	\path [draw, line width=0.7pt, arrows={-latex}] (PNPE) -- (thetaNPE);
    	\path [draw, line width=0.7pt, arrows={-latex}] (thetaNPE) -- (NPE);
    	\path [draw, line width=0.7pt, arrows={-latex}] (NPE) -- (psi);
    	\path [draw, line width=0.7pt, arrows={-latex}] (PMDN) -- (thetaMDN);
    	\path [draw, line width=0.7pt, arrows={-latex}] (psi) -- (MDN);
    	\path [draw, line width=0.7pt, arrows={-latex}] (thetaMDN) -- (MDN);
    	\path [draw, line width=0.7pt, arrows={-latex}] (M) -- (MDN);

    \end{tikzpicture}
  